\documentclass[prl,aps,preprint,superscriptaddress,showpacs,nofootinbib]{revtex4-1}

\usepackage{amsmath}
\usepackage{graphicx}
\usepackage{color}
\usepackage[10pt]{moresize}
\usepackage{amsfonts}
\usepackage{amssymb}
\usepackage{amscd}
\usepackage{enumerate}
\usepackage{epsfig}
\usepackage{subfigure}
\usepackage{graphicx}
\usepackage{bm}
\usepackage{color}
\usepackage{appendix}

\begin{document}

\title{Long-distance free-space measurement-device-independent quantum key distribution}

\author{Yuan Cao}
\thanks{These authors contributed equally to this work}
\affiliation{Hefei National Laboratory for Physical Sciences at the Microscale and Department of Modern Physics, University of Science and Technology of China, Hefei 230026, China.}
\affiliation{Shanghai Branch, CAS Center for Excellence in Quantum Information and Quantum Physics, University of Science and Technology of China, Shanghai 201315, China.}
\affiliation{Shanghai Research Center for Quantum Sciences, Shanghai 201315, China.}

\author{Yu-Huai Li}
\thanks{These authors contributed equally to this work}
\affiliation{Hefei National Laboratory for Physical Sciences at the Microscale and Department of Modern Physics, University of Science and Technology of China, Hefei 230026, China.}
\affiliation{Shanghai Branch, CAS Center for Excellence in Quantum Information and Quantum Physics, University of Science and Technology of China, Shanghai 201315, China.}
\affiliation{Shanghai Research Center for Quantum Sciences, Shanghai 201315, China.}

\author{Kui-Xing Yang}
\thanks{These authors contributed equally to this work}
\affiliation{Hefei National Laboratory for Physical Sciences at the Microscale and Department of Modern Physics, University of Science and Technology of China, Hefei 230026, China.}
\affiliation{Shanghai Branch, CAS Center for Excellence in Quantum Information and Quantum Physics, University of Science and Technology of China, Shanghai 201315, China.}
\affiliation{Shanghai Research Center for Quantum Sciences, Shanghai 201315, China.}

\author{Yang-Fan Jiang}
\affiliation{Hefei National Laboratory for Physical Sciences at the Microscale and Department of Modern Physics, University of Science and Technology of China, Hefei 230026, China.}
\affiliation{Shanghai Branch, CAS Center for Excellence in Quantum Information and Quantum Physics, University of Science and Technology of China, Shanghai 201315, China.}
\affiliation{Shanghai Research Center for Quantum Sciences, Shanghai 201315, China.}

\author{Shuang-Lin Li}
\affiliation{Hefei National Laboratory for Physical Sciences at the Microscale and Department of Modern Physics, University of Science and Technology of China, Hefei 230026, China.}
\affiliation{Shanghai Branch, CAS Center for Excellence in Quantum Information and Quantum Physics, University of Science and Technology of China, Shanghai 201315, China.}
\affiliation{Shanghai Research Center for Quantum Sciences, Shanghai 201315, China.}

\author{Xiao-Long Hu}
\affiliation{State Key Laboratory of Low Dimensional Quantum Physics, Tsinghua University, Beijing, 100084, People's Republic of China.}

\author{Maimaiti Abulizi}
\affiliation{Hefei National Laboratory for Physical Sciences at the Microscale and Department of Modern Physics, University of Science and Technology of China, Hefei 230026, China.}
\affiliation{Shanghai Branch, CAS Center for Excellence in Quantum Information and Quantum Physics, University of Science and Technology of China, Shanghai 201315, China.}
\affiliation{Shanghai Research Center for Quantum Sciences, Shanghai 201315, China.}

\author{Cheng-Long Li}
\affiliation{Hefei National Laboratory for Physical Sciences at the Microscale and Department of Modern Physics, University of Science and Technology of China, Hefei 230026, China.}
\affiliation{Shanghai Branch, CAS Center for Excellence in Quantum Information and Quantum Physics, University of Science and Technology of China, Shanghai 201315, China.}
\affiliation{Shanghai Research Center for Quantum Sciences, Shanghai 201315, China.}

\author{Weijun Zhang}
\affiliation{State Key Laboratory of Functional Materials for Informatics, Shanghai Institute of Microsystem and Information Technology, Chinese Academy of Sciences, Shanghai 200050, People's Republic of China.}

\author{Qi-Chao Sun}
\affiliation{Hefei National Laboratory for Physical Sciences at the Microscale and Department of Modern Physics, University of Science and Technology of China, Hefei 230026, China.}
\affiliation{Shanghai Branch, CAS Center for Excellence in Quantum Information and Quantum Physics, University of Science and Technology of China, Shanghai 201315, China.}
\affiliation{Shanghai Research Center for Quantum Sciences, Shanghai 201315, China.}

\author{Wei-Yue Liu}
\affiliation{Hefei National Laboratory for Physical Sciences at the Microscale and Department of Modern Physics, University of Science and Technology of China, Hefei 230026, China.}
\affiliation{Shanghai Branch, CAS Center for Excellence in Quantum Information and Quantum Physics, University of Science and Technology of China, Shanghai 201315, China.}
\affiliation{Shanghai Research Center for Quantum Sciences, Shanghai 201315, China.}

\author{Xiao Jiang}
\affiliation{Hefei National Laboratory for Physical Sciences at the Microscale and Department of Modern Physics, University of Science and Technology of China, Hefei 230026, China.}
\affiliation{Shanghai Branch, CAS Center for Excellence in Quantum Information and Quantum Physics, University of Science and Technology of China, Shanghai 201315, China.}
\affiliation{Shanghai Research Center for Quantum Sciences, Shanghai 201315, China.}

\author{Sheng-Kai Liao}
\affiliation{Hefei National Laboratory for Physical Sciences at the Microscale and Department of Modern Physics, University of Science and Technology of China, Hefei 230026, China.}
\affiliation{Shanghai Branch, CAS Center for Excellence in Quantum Information and Quantum Physics, University of Science and Technology of China, Shanghai 201315, China.}
\affiliation{Shanghai Research Center for Quantum Sciences, Shanghai 201315, China.}

\author{Ji-Gang Ren}
\affiliation{Hefei National Laboratory for Physical Sciences at the Microscale and Department of Modern Physics, University of Science and Technology of China, Hefei 230026, China.}
\affiliation{Shanghai Branch, CAS Center for Excellence in Quantum Information and Quantum Physics, University of Science and Technology of China, Shanghai 201315, China.}
\affiliation{Shanghai Research Center for Quantum Sciences, Shanghai 201315, China.}

\author{Hao Li}
\affiliation{State Key Laboratory of Functional Materials for Informatics, Shanghai Institute of Microsystem and Information Technology, Chinese Academy of Sciences, Shanghai 200050, People's Republic of China.}

\author{Lixing You}
\affiliation{State Key Laboratory of Functional Materials for Informatics, Shanghai Institute of Microsystem and Information Technology, Chinese Academy of Sciences, Shanghai 200050, People's Republic of China.}

\author{Zhen Wang}
\affiliation{State Key Laboratory of Functional Materials for Informatics, Shanghai Institute of Microsystem and Information Technology, Chinese Academy of Sciences, Shanghai 200050, People's Republic of China.}

\author{Juan Yin}
\affiliation{Hefei National Laboratory for Physical Sciences at the Microscale and Department of Modern Physics, University of Science and Technology of China, Hefei 230026, China.}
\affiliation{Shanghai Branch, CAS Center for Excellence in Quantum Information and Quantum Physics, University of Science and Technology of China, Shanghai 201315, China.}
\affiliation{Shanghai Research Center for Quantum Sciences, Shanghai 201315, China.}

\author{Chao-Yang Lu}
\affiliation{Hefei National Laboratory for Physical Sciences at the Microscale and Department of Modern Physics, University of Science and Technology of China, Hefei 230026, China.}
\affiliation{Shanghai Branch, CAS Center for Excellence in Quantum Information and Quantum Physics, University of Science and Technology of China, Shanghai 201315, China.}
\affiliation{Shanghai Research Center for Quantum Sciences, Shanghai 201315, China.}

\author{Xiang-Bin Wang}
\affiliation{Shanghai Branch, CAS Center for Excellence in Quantum Information and Quantum Physics, University of Science and Technology of China, Shanghai 201315, China.}
\affiliation{State Key Laboratory of Low Dimensional Quantum Physics, Tsinghua University, Beijing, 100084, People's Republic of China.}

\author{Qiang Zhang}
\affiliation{Hefei National Laboratory for Physical Sciences at the Microscale and Department of Modern Physics, University of Science and Technology of China, Hefei 230026, China.}
\affiliation{Shanghai Branch, CAS Center for Excellence in Quantum Information and Quantum Physics, University of Science and Technology of China, Shanghai 201315, China.}
\affiliation{Shanghai Research Center for Quantum Sciences, Shanghai 201315, China.}

\author{Cheng-Zhi Peng}
\affiliation{Hefei National Laboratory for Physical Sciences at the Microscale and Department of Modern Physics, University of Science and Technology of China, Hefei 230026, China.}
\affiliation{Shanghai Branch, CAS Center for Excellence in Quantum Information and Quantum Physics, University of Science and Technology of China, Shanghai 201315, China.}
\affiliation{Shanghai Research Center for Quantum Sciences, Shanghai 201315, China.}

\author{Jian-Wei Pan}
\affiliation{Hefei National Laboratory for Physical Sciences at the Microscale and Department of Modern Physics, University of Science and Technology of China, Hefei 230026, China.}
\affiliation{Shanghai Branch, CAS Center for Excellence in Quantum Information and Quantum Physics, University of Science and Technology of China, Shanghai 201315, China.}
\affiliation{Shanghai Research Center for Quantum Sciences, Shanghai 201315, China.}

\date{\today}

\begin{abstract}
Measurement-device-independent quantum key distribution (MDI-QKD), based on two-photon interference, is immune to all attacks against the detection system and allows a QKD network with untrusted relays.
Since the MDI-QKD protocol was proposed, fibre-based implementations have been rapidly developed towards longer distance, higher key rates, and network verification.
However, owing to the effect of atmospheric turbulence, MDI-QKD over free-space channel remains experimentally challenging.
Here, by developing the robust adaptive optics system, high precision time synchronization and frequency locking between independent photon sources located far apart, we realised the first free-space MDI-QKD over a 19.2-km urban atmospheric channel, which well exceeds the effective atmospheric thickness.
Our experiment takes the first step towards satellite-based MDI-QKD.
Moreover, the technology developed here opens the way to quantum experiments in free space involving long-distance interference of independent single photons.
\end{abstract}
\maketitle

A series of interesting experiments based on Micius satellite have been performed towards global-scale quantum communications \cite{liao2017satellite, Liao2018relay, Yin2017entQKD, yin2017satellite, ren2017satellite}, and fundamental studies in the interface between quantum mechanics and gravity \cite{Xu2019}.
A key next step is to realise quantum interference between independent single photons, also known as the Hong-Ou-Mandel (HOM) interference \cite{mandel1983, paul1986rmp, prasad1987, HOM1987, RMPJWP2012}, after their long-distance travel in free space, which is required by many advanced quantum information tasks such as quantum teleportation \cite{BPMEWZ_97}, entanglement swapping \cite{Pan1998expswapping} and purification \cite{pan2003purification}, and MDI-QKD \cite{LoMDI2012, Braunstein2012sidefree}.
However, the goal remained challenging owing to distortion of the spatial mode and the intensity fluctuation caused by the atmospheric turbulence.

\begin{figure*}[!t]\center
\resizebox{14cm}{!}{\includegraphics{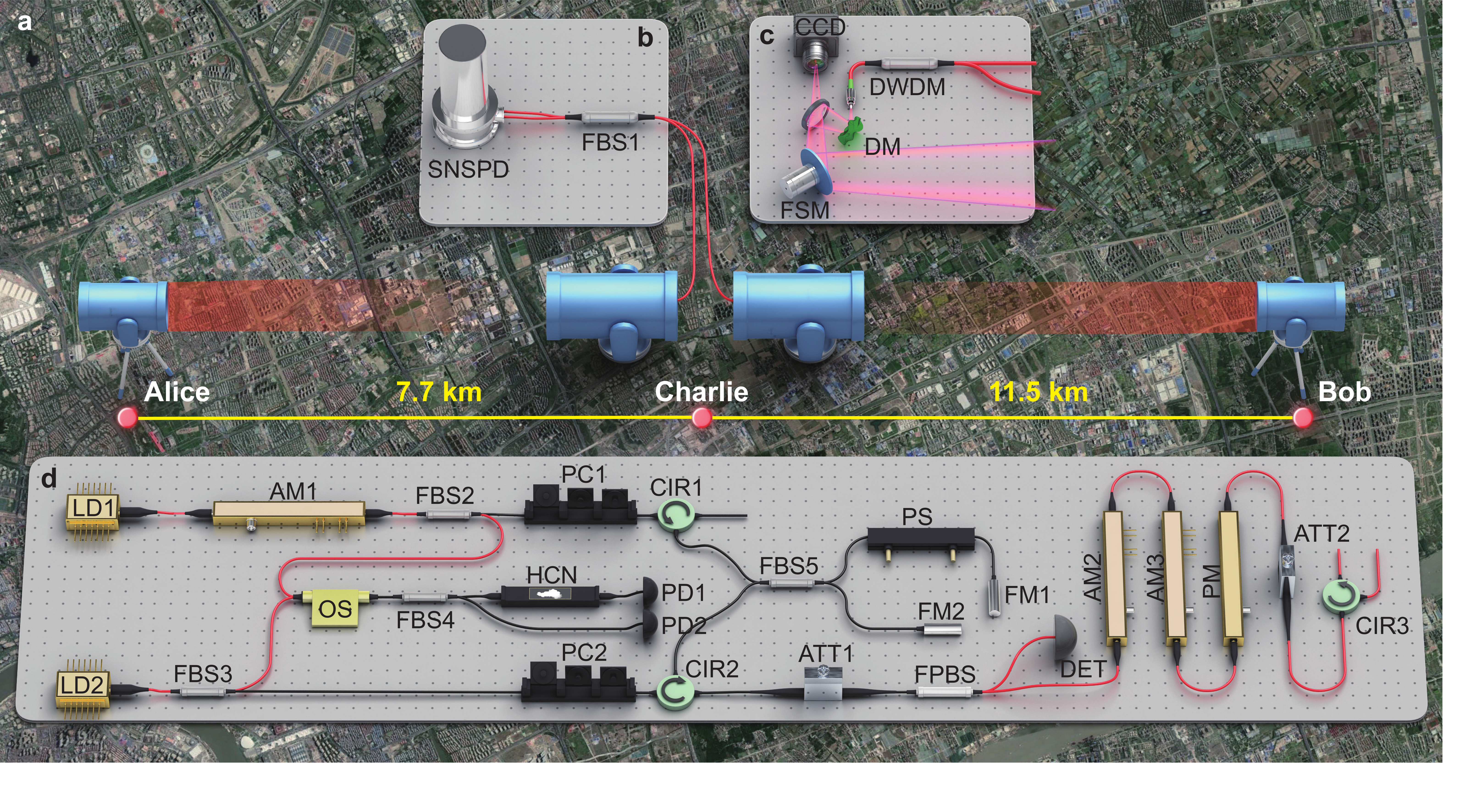}}
\caption{(color online). Setup of free-space MDI-QKD. (a) Top view of the experimental layout at the Pudong area, Shanghai. Alice and Bob are at the opposite direction of the measurement station, with the distance of 7.7 km and 11.5 km, respectively. (b) Arriving photon pulses interference on a fibre beam splitter (FBS) and detected by superconducting nanowire single photon detectors (SNSPDs). (c) A camera and a fast steering mirror (FSM) operate as the APT system. The SPGD algorithm-based AO system compensates the wavefront aberration to maximize the coupling efficiency of SMF by a deformable mirror (DM). (d) The photon sources for MDI-QKD in Alice and Bob. A hydrogen cyanide molecule (HCN) cell is employed to calibrate the wavelength of two DFB laser diodes (LD). The optical pulses generated from LD1 are further modulated by an amplitude modulator (AM) to obtain better uniformity. An asymmetric interferometer is employed to generate pulse pairs. To ensure the relative phase between the two pulses, the interferometer is phase-locked by cw laser emitted from LD2 that has the same wavelength as the signal laser. The encoding is performed by two AMs and a phase modulator (PM). OS, optical switch; CIR, circulator; FM, Faraday mirror; PC, polarisation controller; PS, phase shifter; PD, photodiode; DET, single photon detector.}
\label{Fig:Setup}
\end{figure*}

The theoretical works presented in recent years have shown that two-photon interference from independent photon sources can be used to remove any vulnerability from the detectors in the QKD \cite{LoMDI2012, Braunstein2012sidefree, zhou2016making, hu2018measurement, PhysRevA.87.012320, PhysRevA.85.042307, NCCurty, PhysRevX.9.041012, feihu2019rmp}.
This new form of QKD has been experimentally demonstrated only based on fibre channels \cite{Liu2013expMDI, Rubenok2013expMDI, Tang2014MDI200km, MDI-404km, PirandolaHighrateMDI, ComandarMDIseeded2016, TangMDInetwork2016, LiuHuiPRL}.
Free-space MDI-QKD appeared more challenging because its complex and unstable channel characteristics caused by the atmospheric turbulence.
Here, for the first time, we demonstrate a two-photon interference over a two-link noisy free-space channel from two sites in Shanghai that are separated by approximately 20 km, as shown in Fig.~\ref{Fig:Setup}(a).
We perform a free-space MDI-QKD experiment with a final key rate of 6.11 bps.

As the heart of MDI-QKD, the observation of high-visibility HOM interference requires the indistinguishability of optical pulses that are generated by two independent photon sources and transmitted through two independent free-space channels.
Any mismatch in the degrees of freedom (e.g. spatial, spectral, temporal, and polarisation) lead to a decrement of the visibility.
Normally, however, the free-space optical channels dramatically fluctuates due to the atmospheric turbulence, which is much more difficult to manipulate than fibre-based channels.

The turbulence of atmosphere inevitably affects the wavefront of laser beams and results in the varying random distribution of amplitude and phase at the receiving aperture.
The strength of the turbulence is usually described by the Fried parameter $r_0$ \cite{Fried:66} or Fried's coherence length.
When using receiver telescopes with diameter $D_r$ significantly larger than $r_0$ (typically $\sim 0.5$ to $5$ cm for a 10-km terrestrial free-space channel), the direct interference with two laser pulses propagated through different atmospheric channels suffers from spatial mode distinguishability \cite{LaserRandomMediaBook, Swann:17}.
This can be solved by applying a spatial mode filter, such as single mode fibres (SMFs).
This results in an extra coupling loss, which is typically approximately two orders of magnitude higher than that of using multi-mode fibres.
Therefore, the first challenge is to couple arriving photons into SMFs with an acceptable efficiency.

Adaptive Optics (AO) has been widely used in astronomical observation with large diameter telescopes, where the effect of atmospheric turbulence is the main bottleneck for a further increase in the resolution.
However, the application of AO in a horizontal free-space channel remains uncommon.
One important reason is that the turbulence of horizontal atmosphere in common environment (e.g. urban area) is much stronger than that in vertical atmosphere with good seeing; this issue is challenging for the AO system.
In general, the AO system is composed by a wavefront sensor and a wavefront compensator.
The Shack-Hartmann detector is a good wavefront sensor that can directly estimate the wavefront with a high sampling frequency.
Meanwhile, micro-electro-mechanical system (MEMS)-based or piezoelectric-based deformable mirrors (DM) are common wavefront compensators that contain large number of independent units.
However, under strong turbulence, it is difficult for the wavefront to be effectively estimated owing to intensity scintillations and phase singularities \cite{Barchers:02,Fried:98}.
Therefore, an AO system without wavefront sensors is desired \cite{Vorontsov:97}. Here, we employed a stochastic parallel gradient descent (SPGD) algorithm-based method to perform the wavefront aberration correction \cite{SM}.
With a close-loop bandwidth of $1~kHz$, we performed the AO system to compensate the first 12 orders of Zernike aberration and obtained an improvement in the SMF coupling efficiency by an average of $3$~-~$6~dB$, which implies total improvement of $6$~-~$12~dB$ for two free-space channels.

Besides the spatial mode matching, the timing and frequency mode also require special attentions.
The time duration $\Delta t$ and frequency spread $\Delta \nu$ of interfering photon pulses are required to approach the limitation of the uncertainty principle \cite{RevModPhys.67.759}, $\Delta t\Delta \nu \sim \frac{1}{4\pi}$, to achieve high visibility.
Thus, the time synchronization accuracy $\delta t \ll \Delta t$ and the frequency difference $\delta \nu \ll \Delta \nu$ should satisfy $\delta t \delta \nu \ll \frac{1}{4\pi}$.
In most previous fibre-based experiments that use the interference of two independent photon sources, a master clock was shared between these two sources either by electronic cables or fibres, and strong optical pulses were shared to compare or calibrate frequencies \cite{Tang2014MDI200km}.
However, in the free-space channel, the transmitted optical signals may be disturbed by turbulence, which results in a considerable intensity fluctuation.
Therefore, it becomes a big challenge to share the time and frequency via free-space links.
Here, we employed independent ultra-stable crystal oscillators for each photon source and measurement station \cite{SM}. The standard deviation of the arrival time difference between optical pulses from two sources $\delta t$ is measured to be 32 ps after the feedback control.

Locking a laser cavity on the spectral line of certain atoms or molecules is a widely used technique to obtain stable and narrowly distributed laser frequency.
Acetylene or hydrogen cyanide molecules have many absorption lines in the range of telecom wavelength that are suitable for classical optical communication and quantum communication.
However, the application of this technology to the frequency locking of independent photon sources for two-photon interference has not been experimentally demonstrated.
We employed hydrogen cyanide molecule cells as the frequency standard in each photon source.
As shown in Fig.~\ref{Fig:Setup}(d), laser power is measured before and after the molecule cell by photodiodes to calculate the absorption rate.
With proper temperature control and stable driving current, the frequency difference between two independent distributed feedback (DFB) laser diodes $\delta \nu$ can be limited to 10 MHz.
Fianlly, the good time synchronization and frequency calibration $\delta t\delta \nu \sim 3\times10^{-4}\ll \frac{1}{4\pi}$ enables a large range of selectable $\Delta t$ and $\Delta \nu$.

It is an advantage of using free-space channel that the polarisation maintaining is much easier than in fibre.
The polarisation state can have high fidelity even after the propagation through a near-ground atmosphere on the order of 100 km \cite{Yin:2012,MaXS:2012} or a satellite-to-ground link over 1000 km \cite{liao2017satellite}.
Here, we employed a fibre polarising beam splitter (FPBS) to ensure the indistinguishability of polarisation before interference.
As mentioned above, by tuning the related fibre polarisation controller, approximately 10\% of the arriving photons are reflected by FPBS to directly detect the arrival time for the feedback of time synchronization.

A demonstration of the asymmetric four-intensity decoy-state MDI-QKD protocol \cite{zhou2016making, hu2018measurement, SM} is implemented.
As shown in Fig.~\ref{Fig:Setup}(a), two photon sources are separated by 19.2 km.
The measurement station is placed between them, with distances of 7.7 km and 11.5 km, respectively.
In each photon source, an arbitrarily waveform generator, which is locked on the crystal oscillator, is employed to produce electronic signals for laser diodes and modulators.
The details of the photon source are shown in Fig.~\ref{Fig:Setup}(d).
The optical pulses generated from the DFB laser diodes are further modulated on an amplitude modulator (AM1) to obtain better uniformity and encode decoy states.
An asymmetric Mach-Zehnder (MZ) interferometer is employed to generate coherent pulse pairs that are separated by $\Delta T = 3~ns$.
$Z$-basis encoding is realised by eliminating one of the two pulses, while $X$-basis encoding is realised by applying additional phase between the two pulses.
Thus, the asymmetric MZ interferometer is required to be phase-locked to ensure that the $X$-basis has the same reference frame between different photon sources.
This is done by introducing an additional cw laser to measure the phase difference.
The frequency of the reference cw laser is calibrated by the same molecule absorption cell to ensure that the frequency difference between the cw laser and the signal laser $\Delta \nu_r$ remains below $10~MHz$.
Thus, the accuracy of the estimated phase by the reference cw laser is expected to be better than $2\pi\Delta \nu_r \Delta T\sim0.19~rad$.
Several intensity modulators and a phase modulator are employed to perform the encoding.

\begin{table}\center
\begin{tabular}{ccccccccc}
\hline
\hline
$d_0$ & $E_a^Z$ & $E_a^X$ & $\eta_A$ & $\eta_B$ & $\eta_M$ & $f$ & $\epsilon$ & $N$\\
\hline
$7\times10^{-7}$ & $0.3\%$ & $3\%$ & $17~dB$ & $20~dB$ & $3~dB$ & $1.10$ & $10^{-7}$ & $10^{12}$\\
\hline
\hline
 && $\mu_x$ & $\mu_y$ & $\mu_z$ & $p_o$ & $p_x$ & $p_y$ & $p_z$\\
\hline
\multicolumn{2}{c}{$Alice$} & $0.0394$ & $0.155$ & $0.335$ & $0.0327$ & $0.383$ & $0.0863$ & $0.498$ \\
\hline
\multicolumn{2}{c}{$Bob$} & $0.0713$ & $0.280$ & $0.488$ & $0.0291$ & $0.381$ & $0.0859$ & $0.504$ \\
\hline
\hline
\end{tabular}
\caption{Device parameters for the optimisation of the asymmetric four-intensity decoy-state method \cite{hu2018measurement} for MDI-QKD: dark count rate $d_0$, misalignment error probabilities of $Z$ basis $E_a^Z$ and $X$ basis $E_a^X$, channel efficiency between Alice (Bob) and Charlie $\eta_A$ ($\eta_B$), efficiency of the measurement module in Charlie $\eta_M$, error-correction efficiency $f$, failure probability in the statistical fluctuation analysis of one observable $\epsilon$, and total number of pulse pairs N. The security coefficient of the whole protocol is $\epsilon_{total} = 16\epsilon = 1.6\times10^{-6}$. The optimized source parameters are listed in the lower part of the table.}
\label{tbl:table1}
\end{table}

\begin{figure}[!t]\center
\resizebox{9cm}{!}{\includegraphics{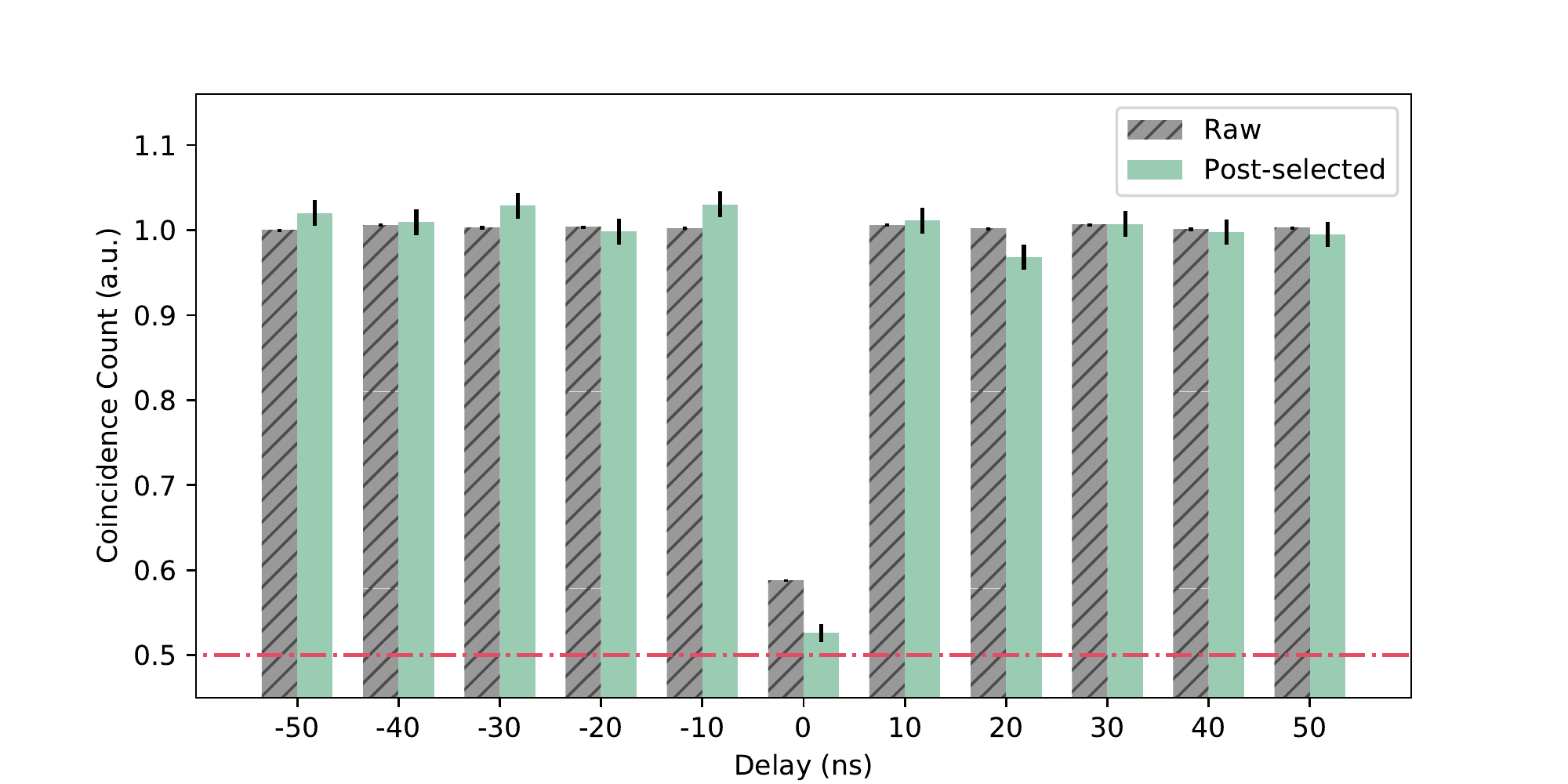}}
\caption{(color online). Coincidence count of the HOM interference over long distance free-space channel. By scanning the time delay of the two detector channels, the coincidence count varies to show the visibility. With all data counted in, the visibility is $0.412\pm0.001$, which is mainly effected by the fluctuation of channel efficiency. Further data post-selecting is taking place with the assistance of an additional reference laser. By ruling out pulse pairs with a different average photon number (by the threshold of 0.98), the visibility of $0.474\pm0.010$ is obtained, which is close to the limitation of HOM interference by coherence states.}
\label{Fig:HOM}
\end{figure}

\begin{figure}[!t]\center
\resizebox{9cm}{!}{\includegraphics{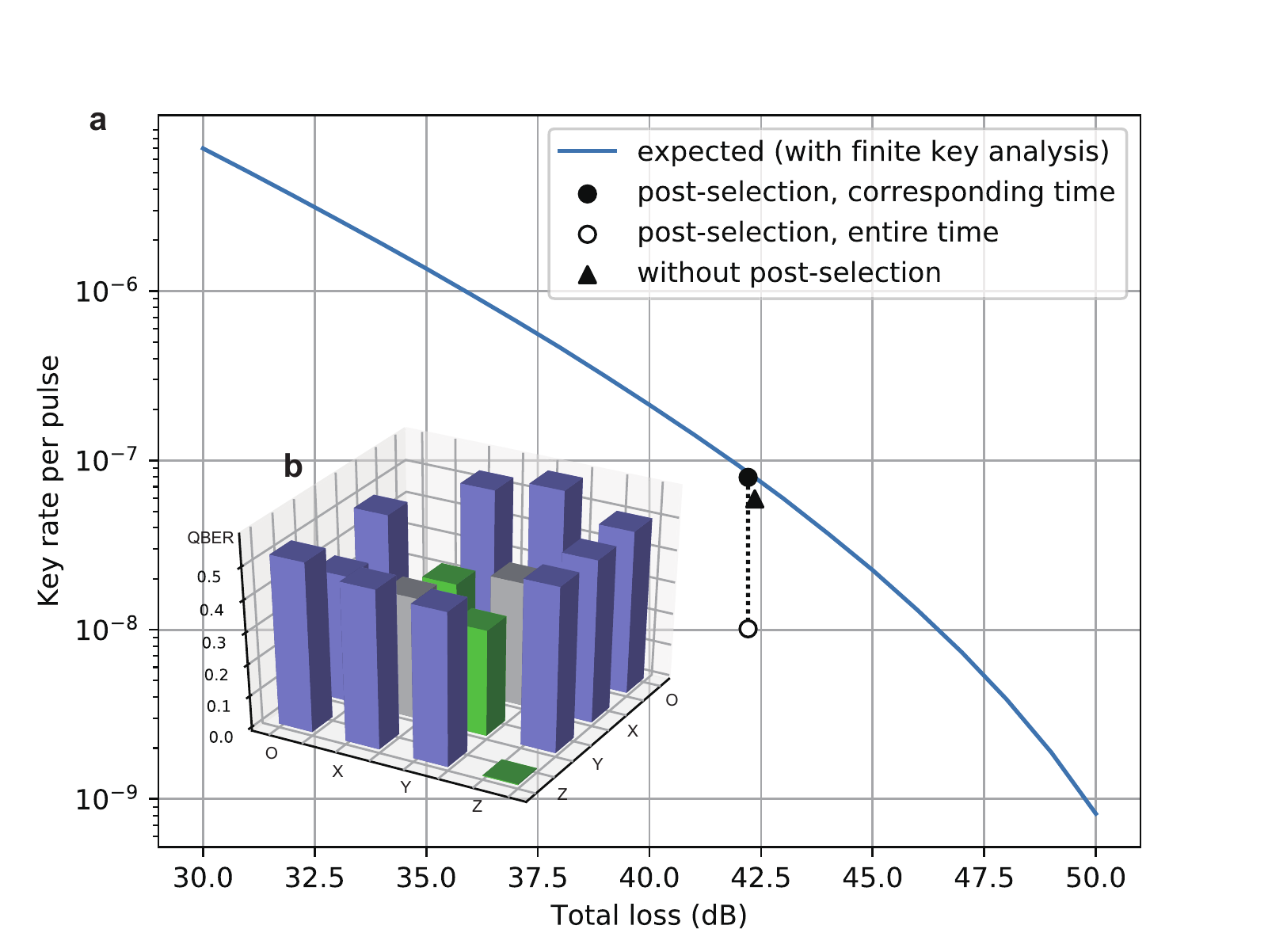}}
\caption{
(color online). Results of MDI-QKD. (\textbf{a}) With the post-selection process, any time slot that have different expected arrival average photon number in $X$-basis is discarded to reduce the QBER in $X$-$X$ basis. $6.15\times 10^6$ time slots, i.e. $6.15\times 10^3$ seconds of valid time, are selected from the data of 13.4 hours by a threshold of 0.8. The final key rate per pulse $7.94\times10^{-8}$ (with the consideration of finite key length) is shown as the solid circle, which is close to the simulated curve. The hollow circle presents the effective secure key rate of post-selected data over the entire data taking duration, i.e. 13.4 hours, which is significantly dropped to $1.01\times10^{-8}$. Secure key rate without post-selection are calculated as $5.93\times10^{-8}$ by data over $6.13\times 10^6$ time slots and shown as the triangle. For a fair comparison, these time slots that consist of four sets of continues acquired data picked from the entire data have similar total loss and valid time with the post-selected data.
(\textbf{b}) Quantum bit error rates (QBER) for each bases. QBERs for $Z$-$Z$ and $X$-$X$ bases are $0.23\%$ and $33.6\%$, respectively. }
\label{Fig:keyrate}
\end{figure}

HOM interference with the visibility of $0.412\pm0.001$ is directly observed on the coincidence count between two superconducting nanowire single photon detectors (SNSPDs), as shown in Fig.~\ref{Fig:HOM}.
The imperfection of visibility is mainly due to the intensity mismatch caused by the atmospheric turbulence.
Together with a fast photodiode, the reference laser for AO can be a good indication to estimate the varying efficiency of free-space channels.
The reference laser powers of every time slot (set as 1 ms) is recorded to perform data post-selection to increase the indistinguishability of the interfering pulses.
For each time slot, the intensity ratio for corresponding pulses is estimated by $min(P_1, P_2)/max(P_1, P_2)$, where $P_1$ and $P_2$ are the power of the reference lasers.
With proper threshold to discard mismatched pulses, the visibility of HOM interference can be increased to $0.474\pm0.010$.
With a set of optimised parameters, as shown in Tab.~\ref{tbl:table1}, over 3.5 M sifted key in the $Z$-$Z$ basis was obtained in 13.4 hours.
The quantum bit error rates for each basis are shown in Fig.~\ref{Fig:keyrate}(b).
Particularly, QBERs for $Z$-$Z$ and $X$-$X$ bases are $0.23\%$ and $33.6\%$, respectively.
With the consideration of finite key length, over 295 kbit of secure key was generated, which corresponded to the key rate of 6.11 bps.
As can be speculated from the results of HOM interference, post-selection can effectively reduce the QBER in $X$-$X$ basis, thus improve the secure key rate over valid time.
The secure key rates with and without post-selection are compared in Fig.~\ref{Fig:keyrate}(a).
Depending on the condition of channel, e.g. the average loss and the strength of atmospheric turbulence, the total amount of secure key after post-selection might decrease owing to the reduction of valid time.
However, under violently turbulent atmosphere, post-selection could be a necessary method to obtain positive final key rate.

In this work, we experimentally demonstrated a 19.2-km free-space MDI-QKD with asymmetric and unstable channels.
The distance achieved in this experiment is well beyond the effective thickness of the aerosphere ($\sim$ 10 km), hence presenting a significant step towards satellite-based MDI-QKD.
The technique developed is also suitable for the fibre-based MDI-QKD system.
In this case, no extra fibre channels are required for the transmission of classical analogue signals such as clock and optical frequency reference.
Therefore, the structure of the MDI-QKD network in free-space and fibre can be significantly simplified.
In the future, a higher clock rate, thus higher key rate, can be obtained by improving the accuracy of time synchronization.
It is worth noting that for shorter pulses, the frequency span will be broadened, which lowers the requirement of optical frequency calibration.
Furthermore, the realisation of two-photon interference in long-distance free-space channels in this work can be applied directly in various other quantum information processing, such as twin-field QKD \cite{lucamarini-twinfield-2018}, quantum teleportation \cite{Teleportation1993}, quantum repeaters \cite{Briegel1998repeater}, and quantum networks \cite{Matthaeus2007Entangling}.

We acknowledge insightful discussions with Teng-Yun Chen, Hao Liang, Fu-Tian Liang, and Li-Hua Sun.
This work was supported by the National Key R$\&$D Program of China (Grants No. 2017YFA0303900, 2017YFA0304000), the National Natural Science Foundation of China (Grants No. U1738201, U1738142, 11654005, 11904358, 61625503, 11822409, and 11674309), the Chinese Academy of Sciences (CAS), Shanghai Municipal Science and Technology Major Project (Grant No.2019SHZDZX01), and Anhui Initiative in Quantum Information Technologies. Y. Cao was supported by the Youth Innovation Promotion Association of CAS (under Grant No. 2018492).

\bibliographystyle{apsrev4-1}

%

\begin{appendices}

\section{S1. Adaptive optics}
The wavefront aberration can be represented by the modal expansion of Zernike polynomials as $\Phi=\displaystyle \sum_{i=1}^{\infty}a_iZ_i$, where $Z_i$ is the Zernike polynomial and $a_i$ is the corresponding coefficient \cite{LaserRandomMediaBook}.
The quantity of $Z_1$ represents a piston, which is the constant change of phase over the entire beam, and it can be ignored here.
$Z_2$ and $Z_3$ represent tilt in the x and y-directions, which is also known as the angle-of-arrival noise and can be suppressed by acquiring, pointing, and tracking (APT) systems \cite{liao2017satellite}.
A tracking camera is installed at the focal plane of the receiving telescope to image the visible beacon laser beam.
Thus, the APT system is able to sense the direction of arriving beam and perform corresponding compensation by a fast steering mirror (FSM), as shown in Fig.~1(b).
The higher order Zernike polynomials represent focus, astigmatism, coma, and so forth, which are expected to be eliminated by the AO technique.

As shown in Fig.~1(c), a 40-unit DM is installed before SMF to perform the wavefront compensation.
A reference laser with the wavelength of 1570 nm is emitted from the transmitter telescope, propagated through the same free-space channel, and coupled by the same SMF with the signal beam.
Thus, the coupling efficiency of the reference beam can be a good indication for the interfering pulses.
After been split by a dense wavelength division multiplexer, the power of the reference beam is measured by an amplified photodiode as the input of the SPGD algorithm.
By adjusting the voltages for each unit of DM, the algorithm attempts to maximise the measured power, thus maximising the SMF coupling efficiency of signal photons.
The dimension of search space for the algorithm can be extended by increasing the number of DM units.
It is worth noting that, under certain strength of turbulence, the coupling efficiency can be considerably improved by correcting the lower orders of Zernike aberration \cite{LaserRandomMediaBook}.

\section{S2. Time synchronization.}
With oven-controlled temperature, the crystal oscillator has a guaranteed short-term stability (Allan Standard Deviation) of $8\times10^{-14}$.
A portion of the arriving photons is directly measured in superconducting nanowire single photon detectors (SNSPD) in the measurement station.
The time and frequency difference of the clocks between the measurement station and each photon source can be estimated and corrected, which results in a 32 ps of standard deviation of the arrival time difference in a long-term.

\section{S3. Asymmetric four-intensity decoy-state MDI-QKD protocol.}
In the situation of free-space MDI-QKD, Alice and Bob have different channel efficiencies and thus have different optimised source parameters.
According to ref. \cite{hu2018measurement}, full optimisation is taken with joint constraints \cite{yu2015statistical,zhou2016making,hu2018measurement}.
In the implementation, each of Alice and Bob has four different intensities: $\{\mu_{ax},\mu_{ay},\mu_{az},o_a=0\}$ for Alice and $\{\mu_{bx},\mu_{by},\mu_{bz},o_b=0\}$ for Bob.
Each time, Alice (Bob) chooses one of these four intensities with the probabilities $p_{ax},p_{ay},p_{az},p_{ao}$ ($p_{bx},p_{by},p_{bz},p_{bo}$) to send a weak coherent state to Charlie, i.e., the measurement station.
The states with the intensities $\mu_{az}$ and $\mu_{bz}$ are prepared in the $Z$-basis, and those with the intensities $\mu_{ax}$, $\mu_{ay}$ $\mu_{bx}$ and $\mu_{by}$ are prepared in the $X$-basis.
The states with the intensity $o_a$ and $o_b$ are the vacuum states.
The key rate of decoy-state MDI-QKD is given by:
\begin{equation}\label{equ:keyrate}
     R = p_{az} p_{bz} \{\mu_{az} \mu_{bz} e^{-(\mu_{az}+\mu_{bz})} \underline{{s_{11}}} [1-H(\overline{e_{11}^{ph}})] - f S_{zz} H(E_{zz})\}
\end{equation}
where $\underline{s_{11}}$ and $\overline{e_{11}^{ph}}$ are the bounds of the counting rate and the phase-flip error rate of single-photon pulse pairs which can be obtained by the decoy-state method \cite{yu2015statistical,zhou2016making,hu2018measurement}, $S_{zz}$ and $E_{zz}$ are the counting rate and the bit-flip error rate when Alice and Bob send pulses with intensities $\mu_{az}$ and $\mu_{bz}$, respectively, $H(p)=-p \log_2 p - (1-p) \log_2 (1-p)$ is the binary entropy function, and $f$ is the correction efficiency.
The four-intensity protocol uses the joint constraints in the statistical fluctuation of different observable \cite{yu2015statistical,zhou2016making,hu2018measurement}, which improves the key rate greatly when the statistical fluctuation is taken into consideration.

\end{appendices}

\end{document}